\documentclass[preprint,aps,floats]{revtex4}

\usepackage{amsmath,amssymb,bm} 
\usepackage[dvips]{color}
\usepackage{graphicx}

\newcommand{\be}{\beta}

\newcommand{\de}{\delta}
\newcommand{\De}{\Delta}

\newcommand{\eps}{\epsilon}

\newcommand{\La}{\Lambda}

\newcommand{\si}{\sigma}

\newcommand{\p}{\partial}
\newcommand{\<}{\langle} 
\renewcommand{\>}{\rangle} 
\newcommand{\txt}{\textstyle}
\newcommand{\dsp}{\displaystyle}
\newcommand\eqn[1]{(\ref{#1})}      
\newcommand{\beq}{\begin{equation}}
\newcommand{\eeq}{\end{equation}}
\newcommand{\ba}{\begin{array}}
\newcommand{\ea}{\end{array}}
\newcommand{\bc}{\begin{center}}
\newcommand{\ec}{\end{center}}
\newcommand{\bea}{\begin{eqnarray}}
\newcommand{\eea}{\end{eqnarray}}
\newcommand{\bi}{\begin{itemize}}  
\newcommand{\ei}{\end{itemize}}
\newcommand{\ben}{\begin{enumerate}} 
\newcommand{\een}{\end{enumerate}}

\newcommand{\ber}{\begin{eqnarray}}
\newcommand{\eer}{\end{eqnarray}}

\newcommand{\half} {{\txt \frac{1}{2}}}

\def\phm{\phantom{-}}

\newcommand{\+}{\dagger}

\newcommand{\up}{\uparrow}
\newcommand{\dn}{\downarrow}
\newcommand{\ah}{{\hat a}}
\newcommand{\bh}{{\hat b}}
\newcommand{\vk}{\bm{k}}

\begin{document}

\title{Mitigating the sign problem 
for non-relativistic fermions on the lattice}
\author{Mark G. Alford}
\affiliation{Physics Department, Washington University,
St.~Louis, MO~63130-4899, USA}
\author{Andrei Kryjevski}
\affiliation{Physics Department,
North Dakota State University,~Fargo, ND~58108-6050, USA }

\begin{abstract}
We study the fermion sign problem
in a theory of non-relativistic fermions
with a spin-independent repulsive interaction.
We work in polar co-ordinates in momentum space, which
makes it straightforward to keep only the low-energy
degrees of freedom close to the Fermi surface.
This is sufficient 
for the purpose of calculating many physically
important low-energy observables.
We find indications that
the sign problem in this
effective theory
will be weaker than in the full theory, so
low-energy properties of the theory could be
calculated by modifying the action to make it positive semi-definite
and including reweighting factors in the observables.
We discuss suitable modifications of the action, and describe 
a possible lattice realization of the
polar momentum space formulation of the theory.
\end{abstract}

\date{10 Nov 2009}

\maketitle
\section{Introduction}
\label{sec:intro}

One of the few known first-principles techniques for calculating the
properties of systems of strongly-interacting fermions is numerical
evaluation of the functional integral.  The standard approach, which
involves Monte Carlo importance sampling of field configurations on a
space-time lattice, has been very successful in zero-density and
high-temperature systems, but at non-zero fermion density and low
temperature it becomes impractical because the fermionic part of the
integrand is not positive and therefore cannot be treated as a weight for
the importance sampling (see, for example,
\cite{Barbour:1997ej,Alford:1998sd,Cox:1999nt,Hands:2007by}).
This has become known as the fermion sign problem.

In this article we investigate the feasibility of 
numerically calculating low-energy
properties of a degenerate Fermi system by focusing on 
fermionic degrees of 
freedom close to the (perturbative) Fermi surface
\cite{hep-ph/0202236,hep-ph/0304156}.  This approach,
known as Landau Fermi liquid effective theory,
allows one to
study low-energy properties of the degenerate Fermi system, such as
heat capacity, density response, and transport coefficients, which are
dominated by the low-energy degrees of freedom 
(see, e.g., \cite{Negele:elastic}). 
We expect this approach to be valid as long as the coupling is
not so strong as to cause major disruption of the Fermi surface, introducing
sensitivity to degrees of freedom far from the perturbative Fermi surface.
As long as this is the case,
the fermion
modes far from the Fermi surface are represented by irrelevant operators 
(see, e.g., \cite{Polchinski:1992ed}).
Some observables that are
sensitive to the degrees of freedom far from the Fermi surface, cannot be
studied in this approach;
these include bulk properties like fermion density and energy density.

We will work in spherical co-ordinates in momentum space. This makes
it straightforward for us to write down the 
leading order terms of the 
Landau Fermi liquid effective theory,
which includes only those degrees of freedom that have momenta
close to the perturbative Fermi surface, and to distinguish between
interactions that change radial momentum $p$ (distance from the Fermi
surface, which is defined by the Fermi momentum $p_F$) 
and those that change the angular component.  The price we
pay is that the interaction term in the effective lagrangian is then
non-local.  It has been argued by Hong and Hsu
\cite{hep-ph/0202236,hep-ph/0304156} that the
corresponding effective theory for QCD 
has a positive semidefinite fermion determinant
to leading order in $1/p_F$, if gluon interactions that 
change a fermion's momentum in the angular direction are suppressed.
The theory we study here is not QCD: it consists of non-relativistic
fermions interacting via an auxiliary scalar field. However, 
it offers a convenient toy model for using the 
Landau Fermi liquid effective theory
to study the sign problem
in dense matter, and it also has physical relevance in
its own right.
We argue that for this theory, without any suppression of
scattering between angular patches,
the effective theory will have a more moderate
sign problem than the full theory has. Then relatively minor changes
to the action (discussed in Sec.~\ref{sec:sign-problem})
will yield a positive-definite action that could be used
to generate an ensemble, and observables in the original theory
could be calculated by reweighting. 
We hope that
future work will investigate this possibility.

Momentum lattice approaches have previously been applied to gauge theories
\cite{Berube:1990ys,Berube:1990ve,hep-lat/9310012,hep-lat/9312050},
but we will restrict our treatment to  non-relativistic fermions that 
feel a spin-independent repulsive interaction, mediated by
an auxiliary bosonic field $A$. 
In Sec.~\ref{sec:fermions} we write down the fermion matrices
in momentum space, in polar co-ordinates, for such a theory,
and  show how the fermion sign problem manifests itself.
In Sec.\ \ref{sec:sign-problem} we discuss ways 
to construct a modified theory
with a positive semidefinite fermion matrix.
In Sec.~\ref{sec:discretization}
we outline a lattice realization of the polar momentum
coordinate formulation of the theory. A more detailed discussion of the
lattice formulation is given in Appendix \ref{app:A-discretization}.

\section{The non-relativistic fermion action}
\label{sec:fermions}

\subsection{Fermions in a background scalar field}

The Euclidean time action for a non-relativistic spin-$\frac{1}{2}$
fermion $\psi$ of mass $m$ in $d$ spatial dimensions, at temperature
$T=1/\be$, interacting
with a background configuration $A(x)$ (where $x=(t,{\bf x}$)) of an
auxiliary real bosonic field, is
\beq
S_{\rm ferm} = \int_{0}^{\beta} \!dt \int \!d^d {\vec x}\,
\psi^{\+}\Bigl(
  \partial_{\tau}-\frac{\nabla^2}{2 m}-\mu+i A(x)\Bigr)\psi 
\label{Lmatrix}
\eeq
The field $\psi$ has spin-up and spin-down components, but the action
is spin-independent so we suppress the spin indices.  
The bosonic
field $A$ has its own action $S_{\rm aux}(A)$ which depends on details
of the interaction between the fermions. 
Its specific form will not concern us
here. If Eq.~\eqn{Lmatrix} were obtained by a
Hubbard-Stratonovich transformation from a purely fermionic
action with a 4-fermion repulsive interaction, then the specific form
of the 4-fermion interaction would determine $S_{\rm aux}(A)$.
In the simplest case the $A$ field could be a discrete spin variable
with $S_{\rm aux}(A)=0$ \cite{Hirsch:1983je}.

We now rewrite this action in terms of a Nambu-Gor'kov-like field
$\Psi$, defined by $\Psi_\up(x) = \psi_\up(x)$, $\Psi_\dn(x) = \psi_\dn^\+(x)$.
The fermion matrix is still diagonal in spin space but no longer
proportional to the identity matrix, so we display it
explicitly:
\beq
S_{\rm ferm} = \int_{0}^{\beta} \!dt \int \!d^d {\vec x}\,
 \Psi^{\+}\left[\begin{array}{cc}
\partial_{\tau}-\frac{\nabla^2}{2 m}-\mu+i A(x)& 0 \\ 0 & \partial_{\tau}+\frac{\nabla^2}{2 m}+\mu-i A(x)\\ 
 \end{array}\right]\Psi
\label{LA0}
\eeq

The partition function of the theory is then
\beq
Z = \int {\rm D} A~\det M_{\rm ferm}(A) \,
  \exp(-{S}_{\rm aux}(A)). 
\label{ZAfull}
\eeq
where $M_{\rm ferm}$ is the fermion matrix, given in  square brackets
in \eqn{LA0}.

Because we assumed the repulsive interaction to be spin-independent,
$M_{\rm ferm}$ can be split into two independent diagonal blocks,
$M_\up$ and $M_\dn$, for spin up and for spin down respectively,
whose determinants are, for general $A(x)$ and $\mu$ (including $\mu=0$)
complex. Therefore the partition function \eqn{ZAfull} has a sign problem,
because the determinant 
of the fermion operator is the square of
a complex number, which is in general not a positive number, and therefore
cannot be used as a Monte-Carlo importance weight.
However, when discretized 
on a spatial lattice of spacing $a$
there is a special value, $\mu_{\rm pos}= d/(ma^2)$, at which this model
has no sign problem because each spin component has a real 
fermion determinant. At $\mu=\mu_{\rm pos}$
the diagonal part of the lattice $\nabla^2$ operator is cancelled, 
leaving only the 
hopping terms. Then, as in the half-filled Hubbard model  \cite{Creutz:1988wv},
there is a particle-hole symmetry, and an associated 
transformation on the fermions that relates $M_\up$
to its complex conjugate, and similarly for $M_\dn$. 


\subsection{Fermions in polar momentum space}
\label{sec:polar-fermions}

We now Fourier transform in space (not in time),
\beq
\Psi(x)=\int_{\bf p}e^{i {\bf p}\cdot {\bf x}} \Psi(t,{\bf p}), \qquad
 A(x)=\int_{\bf p}e^{i {\bf p}\cdot {\bf x}} A(t,{\bf p}),\qquad
 \int_{\bf p}\equiv\int \frac{d^d{\bf p}}{(2 \pi)^d} \ ,
\label{psip}
\eeq
where the momentum-space field has components
$\Psi(t,{\bf p})=(\psi_\up(t,{\bf p}),
 \psi^{\+}_{\downarrow}(t,-{\bf p}))$, obtaining
\beq
\ba{rcr@{}l}
S_{\rm ferm}&=&\dsp \int_{t{\bf p}{\bf q}} \dsp \psi_\up^{\+}(t,{\bf p})
  &\dsp \left(\left[\partial_t+\frac{{\bf p}^2}{2 m}-\mu\right]
  \delta^d({\bf p}-{\bf q})+ i A(t,{\bf p}-{\bf q})\right) 
  \psi_\up(t,{\bf q}) \\[3ex]
&&\dsp {} + \psi_\dn(t,{\bf p})
  &\dsp \left( \left[\partial_t-\frac{{\bf p}^2}{2 m}+\mu\right]
  \delta^d({\bf p}-{\bf q}) -i A(t,-{\bf p}+{\bf q})\right) 
  \psi^{\+}_\dn(t,{\bf q})
\ea
\label{LA0p}
\eeq
Now let us go to spherical coordinates in the spatial momentum. We
write each momentum in terms of a direction $\hat v$ and a
distance $p$ from the perturbative Fermi surface, defined by
$p_F^2=2 m \mu$,
\beq
 {\bf p} = (p_F+p){\hat v},\quad
 \int_{\bf p}=\int_{\hat v}\int_{-\La}^{+\La} 
   (p_F+p)^{d-1} dp \ ,
\label{sph_p}
\eeq
where the factor of $1/(2\pi)^d$ is included in the measure
of the angular integral.
One could think of the
angular variable $\hat v$ as indexing ``patches'' on the Fermi surface.
We have imposed an ultraviolet cutoff $\La\ll p_F$ on $p$,
so only modes close to the Fermi surface are included.
To obtain physical predictions one must integrate out modes with
$|p|>\La$ in the standard way (see, e.g., Ref.~\cite{Polchinski:1992ed}).
Many of the low-energy observables of interest to us, such as transport
properties, vanish in the $T\to 0$ limit and are therefore
insensitive to the U.V. cutoff (recall that there are no
temperature-dependent divergences in a quantum field theory)
as long as $\La$ is much greater than their characteristic momentum scale,
typically of order the thermal momentum $m T/p_F$.

Our momentum-space approach will therefore be valid as long as fermionic 
degrees of
freedom near the Fermi surface play the dominant role in low energy physics.
At sufficiently strong coupling it may become invalid, for example if there
were Bose-Einstein condensation of spatially bound two-fermion states.
Our approach is therefore most relevant to the region of parameter space where
the coupling is too strong to allow perturbative approaches, but
not so strong as to bring in degrees of freedom far from the Fermi surface.
An example would be the onset of superfluidity in a 
moderately-strongly-coupled Nambu--Jona-Lasinio model \cite{Hands:2004uv}.
It is important to note that the momentum-space approach can be used to
map its own region of validity. As long as there is
a range of values of the cutoff $\La$, 
obeying $\La \ll p_F$, where calculated values
of low-energy observables such as transport properties
are independent of the cutoff, we can conclude that fermion modes 
near the Fermi surface are dominating the physics.
If we find that we have to push the 
cutoff up to values of order $p_F$, and there is still $\La$-dependence,
then our assumption has broken down.

The fermion action (\ref{LA0p}) becomes
\beq
S_{\rm ferm}=\int_{t,{\hat u},{\hat v}}\int \!\!dp dq \,
  (p_F+p)^{d-1}(p_F+q)^{d-1}\,
 \Psi^{\+}(t,{\hat u},p)\left[\begin{array}{cc}
M_\up& 0 \\ 0 & M_{\downarrow} \\
 \end{array}\right] \Psi(t,{\hat v},q)
\label{LA0p_sph}
\eeq
where $\Psi(t,{\hat v},q)=(\psi_\up(t,{\hat v},q),\psi^{\+}_{\downarrow}(t,{\hat v},q))$, and
\beq
\ba{rcl}
M_{\up\,p,q,\hat u,\hat v} &=&\dsp
  \left[\partial_t+\frac{p_F}{m}~p+\frac{{p}^2}{2 m}\right]
  \de_{{\hat u},{\hat v}}\de_{p,q}
  +i A(t,p_F({\hat u}-{\hat v})+p{\hat u}-q{\hat v}) \\[3ex]
M_{\dn\,p,q,\hat u,\hat v}&=&\dsp
  \left[\partial_t-\frac{p_F}{m}~p-\frac{{p}^2}{2 m}\right]
  \de_{{\hat u},{\hat v}}\de_{p,q}
  -i A(t,-p_F({\hat u}-{\hat v})-p{\hat u}+q{\hat v}) \ .
\ea
\label{Ms}
\eeq
where $\de_{{\hat u},{\hat v}}\de_{p,q}$ is a simplified notation for
$\delta^{d}({\bf p}-{\bf q})$ written in spherical coordinates.
$M_\up$ describes particles with spin
up, and $M_\dn$ describes holes with spin down. We have explicitly shown 
the indices to make it clear that each is a matrix in momentum space.
The indices are $\hat u$ and $\hat v$, which specify directions in
momentum space,
and $p$ and $q$, which specify the distance from the 
perturbative Fermi surface.

\subsection{Fermion determinant for low-energy modes}
\label{sec:fermdet}

We now focus on the degrees of the freedom near the Fermi surface,
which we expect to be most important for the low-energy properties of
the system. To do that, let us represent (\ref{Ms}) and the
integration measure (\ref{sph_p}) as an expansion in $p/p_F$, where
$p$ is the distance in momentum space from the perturbative Fermi
surface. This will eventually become an expansion in
$m T/p_F^2$ (equivalently $T/\mu$) for observables
that are not sensitive to the cutoff, or in $\Lambda/p_F$
where $\La$ is the ultraviolet cutoff on $p$.
In the rest of this paper we will
focus on the leading order contribution.
The measure then becomes
\beq
 \int_{\bf p}= p_F^{(d-1)} \int_{\hat v}\int_{-\La}^{+\La} dp \ ,
\label{momint}
\eeq
and the fermion matrices are
\ber
&&M_{1\uparrow}=\left[\partial_t+\frac{p_F}{m(\La)}~p\right]{\delta_{{\hat u},{\hat v}}\delta_{p,q}}
  +i A(t,p_F({\hat u}-{\hat v})+p{\hat u}-q{\hat v})\ ,\nonumber \\[2ex]
&&M_{1\downarrow}=\left[\partial_t-\frac{p_F}{m(\La)}~p\right]{\delta_{{\hat u},{\hat v}}
\delta_{p,q}}-i A(t,-p_F({\hat u}-{\hat v})-p{\hat u}+q{\hat v})\ .
\label{M1}
\eer

Eqs.~\eqn{momint},\eqn{M1} are obtained by integrating out
degrees of freedom with $|p|>\La$. This means that irrelevant
couplings (higher-dimension operators, such as higher derivative
terms) are induced, and relevant couplings such as the fermion mass
(and also the parameters of the bosonic field action $S_{\rm aux}$)
must be modified to follow a line of constant physics. 
We have dropped the irrelevant terms since
they are suppressed by powers of 
$\La/\La_0$ where $\La_0$ is the cutoff in the original theory, 
and in \eqn{M1} we have
written the mass as $m(\La)$ as a reminder that $m$ flows with the cutoff.
In the remainder of this paper we
will not show the cutoff-dependence of the relevant parameters
explicitly.  

As noted in Sec.~\ref{sec:polar-fermions}, the momentum space
approach may become invalid if the coupling becomes so strong
that modes far from the Fermi surface start to play an important role.
This will be easily noticed, since it leads to $\La$-dependence
of the low-energy observables.

We expect the effective theory \eqn{M1} to have a weaker
sign problem than the full theory \eqn{LA0} 
because it is closer to having a particle-hole symmetry:
we have discarded the degrees of freedom far from the
Fermi surface, which, in the full theory 
have very different phase space and dispersion relations above and 
below the Fermi surface.
As we will see below, in the effective theory there is a
variable transformation that almost guarantees the
positivity of the fermion determinant: it is only violated
by the $p{\hat u}-q{\hat v}$ piece in the interaction term.
In Sec.~\ref{sec:sign-problem} we will discuss how this
piece might affect the fermion determinant;
it is formally subleading in $p/p_F$, but
for now we keep it in the action because we do not know {\it a priori}
the typical scale of variation of the relevant A-modes,
so it may not be rigorously negligible.

We expect low-energy degrees of freedom to be well described by
the leading-order fermion matrices (\ref{M1}). This theory still
has a sign problem: 
$\det M_{\rm ferm} = \det M_{1\up}\det M_{1\dn}$ is in general
complex. To see why, consider the complex conjugate of $\det(M_{1\up})$:
\beq
\ba{rcl}
\det(M_{1\up})^* &=& \det( M^*_{1\up} ) \\
&=& \det\Bigl( (\partial_t+\frac{p_F}{m}~p)\de_{{\hat u},{\hat v}}
\delta_{p,q} -i A^*(t,p_F({\hat u}-{\hat v})+p{\hat u}-q{\hat v}) \Bigr)\\
&=& \det\Bigl( (\partial_t+\frac{p_F}{m}~p)\de_{{\hat u},{\hat v}}
\delta_{p,q} -i A(t,-p_F({\hat u}-{\hat v})-p{\hat u}+q{\hat v}) \Bigr)\\
&=& \det\Bigl( (\partial_t-\frac{p_F}{m}~p')\de_{{\hat u},{\hat v}}
\delta_{p',q'} -i A(t,-p_F({\hat u}-{\hat v})+p'{\hat u}-q'{\hat v}) \Bigr)\ . 
\ea
\label{det-Mup}
\eeq
where we have used the fact that the auxiliary field $A$ is
real, so $A^*(t,{\bf p})=A(t,-{\bf p})$, and  
that the relabelling $p'=-p$, $q'=-q$ does not
change the determinant. 
Compare this with the determinant for the spin-down particles
\beq
\det(M_{1\dn}) = \det\biggl(
  (\partial_t-\frac{p_F}{m}~p){\de_{{\hat u},{\hat v}}\delta_{p,q}}
  -i A(t,-p_F({\hat u}-{\hat v})-p{\hat u}+q{\hat v}) \biggr)\ .
\label{det-Mdn}
\eeq
For a general $A$-field configuration, $\det(M_{1\up})^*$
differs from $\det(M_{1\dn})$, so the total weight 
$\det(M_{1\up})\det(M_{1\dn})$ will not in general
be positive semidefinite.
However we can see that $\det(M_{1\up})^*$ is some sense ``almost''
equal to  $\det(M_{1\dn})$. 
Only the last term in the argument of the $A$ field 
differs between \eqn{det-Mdn} and the last line of \eqn{det-Mup}.
If the offending terms could be ignored or suitably modified 
(for example if the $p/p_F$ expansion converged well enough,
or if scattering between patches proved to be negligible---see below) 
then the fermion matrix would be positive-semidefinite.

\section{Modified effective action without a sign problem}
\label{sec:sign-problem}

To test the idea that the effective theory \eqn{M1} has a
tractably weak sign problem, one must modify it to obtain
an exactly positive-semidefinite weighting, and calculate
observables with a reweighting factor that gives results appropriate
to the action \eqn{M1}. If the sign problem is sufficiently weak,
the reweighting factor will not be too noisy, and useful results
can be obtained.
One approach would be
to generate ensembles weighted by $|\det M_1|$ or $|{\rm Re}\,\det M_1|$,
and then reweight back to the proper $\det M_1$ weighting.
This has not yet been attempted, perhaps because in the
position-space formulation \cite{hep-ph/0202236,hep-ph/0304156}
the calculation of $\det M_1$ requires the construction of a non-local operator
which is used in the coupling of the fermion modes to the auxiliary field.
We suggest that formulating the theory in momentum space, as we have
done in this paper, may make it easier to perform numerical investigations
of the high density effective theory. For example, it is easy to
do small-scale studies using a small number of angular patches on the
Fermi surface.

In the remainder of this section we will describe an alternative
to using $|\det M_1|$ or $|{\rm Re}\,\det M_1|$, namely
modifying the interaction term in (\ref{M1}) so as to
make the fermion determinant positive-semidefinite.

\subsection{Action with no scattering between patches}
\label{sec:HH}

The simplest and most radical approach would be to make the different
directions on the Fermi surface
independent of each other, by coupling fermion modes only to
auxiliary field modes that scatter them within
the same angular patch on the Fermi surface.

To see that such a theory is positive semidefinite, consider a 
theory in one spatial dimension ($d=1$)
where the angular variable
$\hat u$ can take on only two values, $\hat a$ and $-\hat a$,
and we discard high-momentum $A$-field modes that could scatter
fermions from $\ah$ to $-\ah$.
This situation was studied in Sec.~II of Ref.~\cite{hep-ph/0304156}
(see also \cite{Dougall:2007fp}).
The fermion action for a single spin state is
\beq 
\ba{rl}
S^{(\up)}_{\rm ferm} = &\dsp 
\int dp dq \, \psi^{\+}(\ah,p)\Bigl( \de(p-q)\p_t + i A((p-q)\ah) 
  + \frac{p_F}{m} p \de(p-q)\Bigr) \psi(\ah,q) \\[2ex]
+ &\dsp
\int dp dq \, \psi^{\+}(-\ah,p)\Bigl( \de(p-q)\p_t + i A((p-q)(-\ah)) 
  + \frac{p_F}{m} p \de(p-q)\Bigr) \psi(-\ah,q) \ .
\ea
\label{our1+1D}
\eeq
In each line, the first two terms are anti-Hermitian 
and the last one is Hermitian.
Now, in the $-\ah$ patch (second line of \eqn{our1+1D})
we change variables to $p'=-p$ and $q'=-q$ and then rename these
back to $p$ and $q$, giving
\beq
\ba{rl}
S^{(\up)}_{\rm ferm} = &\dsp 
\int dp dq \, \psi^{\+}(\ah,p)\Bigl( \de(p-q)\p_t + i A((p-q)\ah) 
  + p \de(p-q)\Bigr) \psi(\ah,q) \\[2ex]
+ &\dsp
\int dp dq \, \psi^{\+}(-\ah,-p)\Bigl( \de(p-q)\p_t + i A((p-q)\ah) 
  - p \de(p-q)\Bigr) \psi(-\ah,-q) \ .
\ea
\label{our1+1D-inv}
\eeq
Note that the fermion modes living at $-\ah$ are different from the modes
at $\ah$, but they couple to the same auxiliary field modes.
Equation \eqn{our1+1D-inv} leads to a fermion matrix of the form
\beq
M^{(\up)} = \left(
\ba{cc} M_A+M_H & 0 \\
 0 & M_A-M_H
\ea\right)
\label{blocks}
\eeq
where $M_H$ is a Hermitian operator ($p$ in this case)
and $M_A$ is an anti-Hermitian operator ($\p_t+iA$ in this case).
We assume that both are of even dimension.
It then follows that $M^{(\up)}$ has positive-semidefinite determinant, 
since the eigenvalues of $M_A-M_H$ are (up to a sign) the complex
conjugates of the eigenvalues of $M_A+M_H$. The spin-down matrix
$M^{(\dn)}$ is independently positive-semidefinite, via a similar argument.
This agrees with the result obtained in Ref.~\cite{hep-ph/0304156}.

It was important in obtaining this result that the 
off-diagonal 
blocks in \eqn{blocks} were zero: in a $d=1$ theory this
restriction corresponds to putting a cutoff $\La_A$ on 
the auxiliary field modes, where $\La \ll \La_A \ll p_F$,
which imposes the condition that the $A$-field can only scatter fermions
{\em within} angular patches.
In higher dimensions, however, there is no separation 
of scales between
momenta that scatter between patches and momenta that scatter
within patches: an $A$-field of arbitrarily small momentum
can scatter a fermion from one patch to a neighboring one,
and the fermion matrices are no longer positive semidefinite 
(see Sec.~\ref{sec:tests}).
The argument made in Ref.~\cite{hep-ph/0304156} is that,
in QCD, even in more than one dimension, 
scattering between neighboring patches is subleading in
$\mu R$ (where $R$ is the in-medium screening distance for gluons)
so scattering between patches can be ignored, 
and high-density QCD can be treated as a collection
of $d=1$ theories.
However, in a first
principles non-perturbative approach one cannot rely on such
essentially perturbative results. The screening distance is {\it a priori}
unknown and should be determined in a simulation. Therefore, in this
article we will not assume that scattering between patches can be
neglected.


\subsection{Action that is symmetrized in the residual momentum}
\label{sec:sym}

We propose an alternative way of making the fermion determinant
positive semidefinite: modify the coupling to the background field
by symmetrizing it in the residual momentum $p$.
This means using a new fermion determinant $M_{1s}$ which
is constructed by 
replacing the background field $A$ in \eqn{M1} with
\beq
A_{s}({\bf p}-{\bf q})=\frac{1}{2}\Bigl(
  A(p_F({\hat u}-{\hat v})+p{\hat u}-q{\hat v})
 +A(p_F({\hat u}-{\hat v})-p{\hat u}+q{\hat v}) \Bigr).
\label{Asym}
\eeq
The fermion determinant with this symmetrized coupling to
the auxiliary field is guaranteed to be non-negative 
for any $A({\bf p})$. 
This follows from the argument of section \ref{sec:fermdet}:
by coupling to $A_s$ we make the interaction term symmetric
in the residual momenta $p$ and $q$, so now 
$\det M_{1s\dn} = (\det M_{1s\up})^*$, and
the symmetrized full determinant
$\det M_{{\rm ferm},s} = \det M_{1s\up}\det M_{1s\dn}$ is real and
positive for any $A$-field configuration.

The modified ensemble generated using $A_s$ has larger violation
of momentum conservation (see Appendix~\ref{app:A-discretization})
which will be corrected by reweighting back to
the original fermion matrix (see Sec.~\ref{sec:conclusions}).
Since the modification is (formally) subleading in
$p/p_F$ and hence $\La/p_F$, the reweighting factor may be sufficiently
well-behaved that the modified theory is still suitable 
for calculations of low-energy observables in the original theory.
That is one of the main conjectures of this paper, and we hope
it can be tested by explicit Monte-Carlo calculations.

A more radical modification of the fermion action
would be to retain only the leading piece in the expansion, 
$A(p_F({\hat u}-{\hat v}))$, throwing out all dependence on
the residual momenta. This would correspond to having all fermion modes
in a given patch $\hat u$ couple to all modes in another patch $\hat v$
via the {\em same} auxiliary field mode.
This might serve as
a good starting point in numerical investigations because of its
relative simplicity.

It needs to be emphasized that, unfortunately, the proposed
approximation may not be systematically improved. The approximation
hinges on the observation that the
theory with couplings symmetrized in $p$ (which corresponds to
discarding terms in its Taylor expansion that are odd powers of $p$)
is positive semidefinite. 
Inclusion of those odd powers and other higher-order terms in
$p/p_F$ will inevitably re-introduce the complexity existing in the
unmodified theory (\ref{M1}), even more so in (\ref{LA0p_sph}),(\ref{Ms}). 
However, our conjecture is that simulations
based on the modified theory described above will capture low-energy
properties of the full theory with a controllable error of order of
$T/\mu$.

\section{Lattice realization of the theory}
\label{sec:discretization}

\begin{figure}
\includegraphics[scale=0.6]{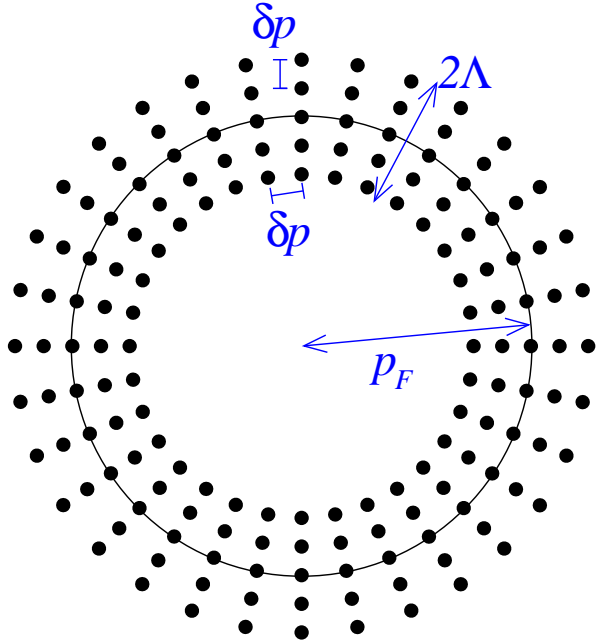}
\caption{
Discretization of the space of fermionic momentum modes near a
two-dimensional Fermi surface. This example has
momentum lattice spacing 
(infrared cutoff) $\de p=p_F/8$, and ultraviolet
cutoff $\La=p_F/4$.
}
\label{fig:discretization}
\end{figure}

To numerically evaluate expectation values of observables
by Monte-Carlo methods, it is necessary to
formulate the theory on a lattice. We will discretize the Euclidean
time direction in the standard way, with a lattice of $N_t$ points
covering the range $[0,1/T]$.
The spatial directions will be discretized as a lattice
in momentum space. As described above we use polar co-ordinates: we
discretize the angular direction $\hat u$ and the radial direction $p$,
and keep only modes near the Fermi surface by imposing an
ultraviolet (UV) cutoff $\La$ on the residual momentum $p$.
We choose a single infrared (IR) cutoff $\de p$, which
is the momentum lattice spacing in the radial direction, and we
choose this to also be
the momentum lattice spacing in the angular direction on the inside
edge of the momentum shell (at $p=-\La$).
An example of such a momentum lattice for fermions in two dimension is shown in
Fig.~\ref{fig:discretization}.
In order to simulate properties of infinite volume continuum
theory the lattice parameters need to satisfy
\beq
\La \gg p_{\rm typ} \gg \de p \ ,
\label{lat_ineq}
\eeq
where $p_{\rm typ}$ is the characteristic momentum scale of the physics
of interest, for example the thermal momentum $m T/p_F$, or a 
momentum carried by a correlation function. 


The modes of the auxiliary field $A$ are
most conveniently chosen to lie on a cubic lattice
of lattice spacing $\de p$ in all spatial directions. In order to
couple them to fermion modes that live on a polar momentum
lattice, while maintaining the correct
large-volume ($\de p\to 0$) limit, we must allow a small violation
of conservation of momentum, which goes to zero as $\de p\to 0$.
This arises because, in coupling the auxiliary field to the
fermions, we have to treat momenta that differ from each
other by an amount less than half the momentum resolution $\de p$
as the same momentum.
Then a pair of Fermion states labeled by ${\hat u},~p$ and ${\hat v},~q$
corresponding to a momentum difference
$\Delta{\bf p}=p_F({\hat u}-{\hat v})+p{\hat u}-q{\hat v}$,
will be coupled to all cubic lattice modes $A({\bf k})$
such that $|\Delta{\bf p}-{\bf k}|<\de p/2$.

Momentum violation at the scale of an infrared cutoff is not unusual:
it arises when one performs lattice calculations
in a box with hard-wall or reflecting boundary conditions, in which
case momentum
conservation is violated at distance scales on the order of the size
of the box, but this artifact disappears in the large-volume limit.
For a more detailed discussion, see appendix \ref{app:A-discretization}.
From \eqn{M1}, our discretized fermion matrices are
\beq
\ba{rcl}
M^{\rm latt}_{1\up}&=&\dsp
  \biggl(\partial_t+\frac{p_F}{m}~p\biggr){\de_{{\hat u},{\hat v}}\de_{p,q}}
  +i \sum_{\vec k} A(t,\vec k)
    W(p_F({\hat u}-{\hat v})+p{\hat u}-q{\hat v}-\vec k)\\[3ex]
M^{\rm latt}_{1\dn}&=&\dsp
  \biggl(\partial_t-\frac{p_F}{m}~p\biggr){\de_{{\hat u},{\hat v}}\de_{p,q}}
  -i \sum_{\vec k}  A(t,-\vec k)
    W(-p_F({\hat u}-{\hat v})-p{\hat u}+q{\hat v}-\vec k) \ .
\ea
\label{M1disc}
\eeq
In these expressions, $W(\vec q\/)$ (see \eqn{Mint-disc}) is
a function that imposes approximate momentum
conservation. It is is peaked around $|\vec q\/|=0$, with 
a width $\approx \de p$.
The role of the $W$ function is to define a mapping
between the $A$-field momenta $\vec k$, which lie on a cubic lattice
of spacing $\de p$, and the polar lattice momentum differences
$\Delta \vec p=p_F({\hat u}-{\hat v})+p{\hat u}-q{\hat v}$, such that
$|\vec k-{\Delta \vec p}|<\de p/2$.

As in the momentum continuum 
(Sec.~\ref{sec:sign-problem}), the lattice fermion operators \eqn{M1disc} have 
non-positive determinants, but one can modify the coupling
to the background field
so that a given pair of fermion fields
$\psi^\dagger(\hat u,p)\psi(\hat v,q)$
couples to the symmetrized combination of auxiliary fields
$\half\{A(p_F(\hat u-\hat v)+p\hat u - q \hat v)+
A(p_F(\hat u-\hat v)-p\hat u + q \hat v)\}$. 

The result is a pattern of coupling between the $A$-field and the fermions
that has the fermion matrices $M_{1s}$, where
\beq
\ba{rcl@{}l}
M^{\rm latt}_{1s\up}&=&\dsp
 (\partial_t+\frac{p_F}{m}~p){\de_{{\hat u},{\hat v}}\de_{p,q}}
  +\frac{i}{2} \sum_{\vec k} A(t,\vec k) \Bigl( &\dsp
   \phm\, W(p_F({\hat u}-{\hat v})+p\hat u - q \hat v -\vec k) \\
 &&&\dsp +\, W(p_F({\hat u}-{\hat v})-p\hat u + q \hat v -\vec k) \Bigr) \\[3ex]
M^{\rm latt}_{1s\dn}&=&\dsp
  (\partial_t-\frac{p_F}{m}~p){\de_{{\hat u},{\hat v}}\de_{p,q}}
  -\frac{i}{2} \sum_{\vec k}  A(t,\vec k) \Bigl( &\dsp
   \phm\, W(-p_F({\hat u}-{\hat v})-p{\hat u}+q{\hat v}-\vec k) \\
 &&&\dsp +\, W(-p_F({\hat u}-{\hat v})+p\hat u - q \hat v -\vec k)
 \Bigr)
\ea
\label{M1s}
\eeq

It is then straightforward to verify, following the same procedure as
in Eqns.~\eqn{det-Mup} and \eqn{det-Mdn}, that 
$\det(M_{1s\up})^* = \det(M_{1s\dn})$, so the 
fermion matrices \eqn{M1s} have no sign problem, and their
determinants can be used as Monte-Carlo weights to generate
an ensemble of configurations. As noted in Appendix~\ref{app:A-discretization},
in the modified theory an auxiliary field mode with momentum $k$
couples not only to pairs of fermion modes  whose momentum difference
is within $\de p$ of $k$, but also to pairs of fermion modes with
momentum differences that deviate from $k$ by as much as $\La$.
This additional momentum
non-conservation should be corrected by the reweighting back to
the original fermion matrix (see Sec.~\ref{sec:conclusions}), 
and it remains to be seen whether it will
lead to a significant sign problem in computing the reweighting factor.

\section{Conclusions}
\label{sec:conclusions}

We have formulated a theory of non-relativistic fermions
with a spin-independent repulsive interaction in polar coordinates
in momentum space. This makes it very straightforward for us
to write down the high-density effective theory for
the fermionic modes with the lowest free energy 
(those near the perturbative Fermi 
surface) which are expected to be relevant for low-energy properties of 
the system.
We have shown that in general this
effective theory still has a sign problem. We argued that
various modifications of the action 
would yield a positive semidefinite weighting factor  $W_{\rm pos}$
which is suitable for Monte-Carlo sampling.
These include completely decoupling different angular patches to give
a set of decoupled 1+1-dimensional theories 
(Sec.~\ref{sec:HH} and \cite{hep-ph/0304156});
symmetrizing the interaction in the residual momentum $p$,
i.e.~the momentum distance from the Fermi surface (Sec.~\ref{sec:sym});
and taking the modulus of the fermion determinant. The last two
modifications are (formally) subleading in $p/p_F$.

To obtain physical results one would have to re-weight
back to the original theory, by calculating the reweighting
factor ${\cal R}=\< \det M_1/W_{\rm pos}\>_{\rm pos},$ where 
$M_1$  is the fermion determinant for the unmodified effective
theory of the low-energy degrees of freedom \eqn{M1}, and
$\<\cdots\>_{\rm pos}$ is an average in the ensemble weighted by
$W_{\rm pos}$. If ${\cal R}$ turns out to have sufficiently
small fluctuations on lattices whose volume is
big enough to be physically relevant, then the reweighting procedure
can be used to calculate observables in the original theory.
A priori, we cannot say which of the modifications proposed here
will yield a positive ensemble with the closest overlap with the physical
ensemble (i.e.~the smallest fluctuations in ${\cal R}$), or whether
any of them will be close enough to be useful. This is a typical
issue with reweighting schemes. One may get some information
by comparing simulations that use different modifications.
Our approach reduces the original theory to a Yukawa-type theory
of the degrees of freedom near the Fermi surface, and it is
encouraging that
such theories have been found to have a fairly gentle sign problem,
so that simulating using the modulus of the fermion determinant
did not introduce unphysical phases \cite{Hands:2004uv}.
Rough preliminary investigations of a two-dimensional model
(see Appendix~\ref{app:testing})
seem to indicate that in the momentum space formulation
the sign problem is indeed very moderate.
The momentum space
approach may become invalid if the coupling becomes so strong
that modes far from the Fermi surface start to play an important role.
As we noted in Sec.~\ref{sec:polar-fermions}, it will be straightforward
to check whether this is a problem, since such a breakdown will
be signaled by $\La$-dependence
of the low-energy physical observables.

Working in momentum space gives the theory a perfect kinetic term.
We keep only the leading order in $p/p_F$ but
there are no discretization errors coming from
approximating derivatives by finite spatial differences.
The polar momentum formulation also makes it
straightforward to develop simplified models 
for testing the sign problem in the
effective theory,
such as ones with an auxiliary field that takes on discrete
values (see Appendix~\ref{app:testing}) or with
a small number of patches on the Fermi surface
(see Sec.~\ref{sec:1patch}).
However, it imposes some costs.
Firstly, as discussed in Appendix~\ref{app:A-discretization}, we have
to introduce an infrared cutoff $\de p$, and in the interaction
between the auxiliary field and the fermions we treat fermion modes
whose momenta that are within $\half\de p$ of each other as if they
had the same momentum. This means the interactions introduce
momentum violation of order $\de p$.  Secondly, our formulation is
more expensive for lattice computations because
the interaction term, which was local in position space,
becomes non-local in momentum space. This means that
$M_1$ and any modified version of it are non-sparse matrices.
(Note, however, that the position-space formulation 
\cite{hep-ph/0202236,hep-ph/0304156} also requires
the calculation of a non-local operator).
Computation of the determinant of
a generic non-sparse $N\times N$ matrix takes of order $N^3$ operations
\cite{Watkins2002}, whereas the algorithms for dynamical
fermions with local actions typically take of order $N$
operations at fixed UV cutoff \cite{arXiv:0810.5634}. 
This extra cost will easily be worthwhile, however, if the
resultant ensemble has a less severe sign problem.
The number of configurations required for reweighting from
a positive action to the original action
rises extremely quickly, as $\exp(N \De f)$ 
(where $\De f$ is the free energy density difference between the
actions \cite{Cox:1999nt}), so if
$W_{\rm pos}$ has a smaller value of $\De f$
than existing approaches then this
could easily compensate for the extra costs arising from
the non-sparse fermion matrix. We hope that lattice
calculations using the high-density effective theory
will test these ideas in the future.

\section*{Acknowledgements}
We thank Claude Bernard, Philippe de Forcrand, Deog-Ki Hong, 
Steve Hsu, John Laiho, and Dean Lee
for discussions. This research was
supported in part by the Offices of Nuclear Physics and High
Energy Physics of the U.S.~Department of Energy under contracts
\#DE-FG02-91ER40628,  
\#DE-FG02-05ER41375, 
\#DE-FG52-08NA28921. 

\newpage

\appendix  

\section{Discretization in polar momentum space}
\label{app:A-discretization}

In this appendix we discuss some of the technical issues
that arise when we discretize the fermion degrees
of freedom in polar coordinates in momentum space.
The main question is how the fermions should be coupled
to the auxiliary field, which is also discretized in momentum space.

\subsection{Entire Fermi surface}
\label{sec:whole-surface}

\begin{figure}[h]
\includegraphics[scale=0.5]{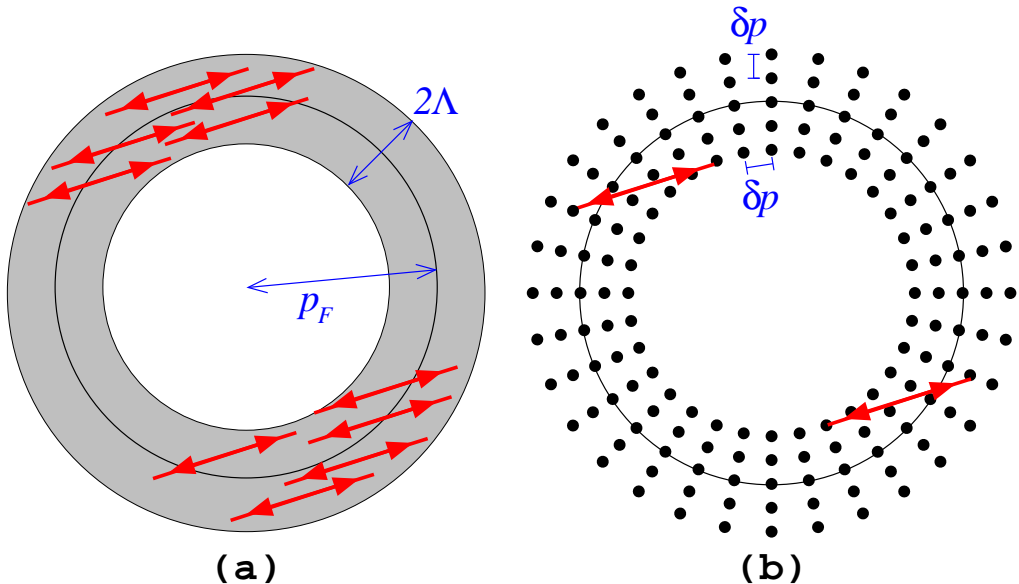}
\caption{Momentum-conserving
coupling of auxiliary field $A$ to modes near the 
Fermi surface, in two spatial dimensions.
Panel (a): continuous momentum space, where an $A$-mode with given momentum 
will couple to any
pair of fermions with the appropriate momentum difference
(thick lines with embedded arrow heads, red online; just a few
examples are shown).
Panel (b): discretized momentum space for fermions
using polar co-ordinates. If we require exact momentum conservation,
then a given (non-radial) $A$-mode can couple 
to only one pair of fermion modes and their antipodal images.
}
\label{fig:continuum}
\end{figure}

If we discretize the whole Fermi surface then we cannot
maintain exact momentum conservation in the interactions
that are mediated by the auxiliary $A$ field.
We illustrate the issue for the two-dimensional case
in Fig.~\ref{fig:continuum}.
In the infinite volume limit (continuum in momentum space)
the $A$-field mode with a given momentum $\vec k$
can couple to any two fermion modes whose momenta differ by $\vec k$:
the Lagrangian will contain many terms of the form 
$\psi^\+(\vec p)A(\vec k)\psi(\vec p-\vec k)$
for different values of $\vec p$. These pairs of modes densely cover
the available momentum space, as shown 
in Fig.~\ref{fig:continuum}a by the many ways one can draw lines
of momentum $\vec k$ (thick lines with embedded arrow heads)
connecting pairs of fermion modes near the Fermi surface.
However, once one discretizes the radial and angular components of
the momenta, the momentum difference between any two fermion modes
becomes unique (up to inversion through the origin, and
apart from purely radial momentum differences).
So in the discretized theory (Fig.~\ref{fig:continuum}b)
if we require exact conservation of momentum then
a typical $A$-mode either cannot couple to any fermion modes, or can 
only couple to one pair of fermion modes, and their antipodal 
images. This remains true no matter how small $\de p$ becomes.
We conclude that an exactly-momentum-conserving discretization
will not give the correct theory in the $\de p\to 0$ limit.
For example, if one made the coupling constant $g$ small,
most fermion-fermion scattering amplitudes would 
be proportional to $g$ in the full theory (from
single $A$ exchange) but would have no ${\cal O}(g)$
contribution in the discretized theory, where single $A$ exchange
would be forbidden by momentum conservation.

\begin{figure}
\parbox{0.4\hsize}{
\includegraphics[scale=0.5]{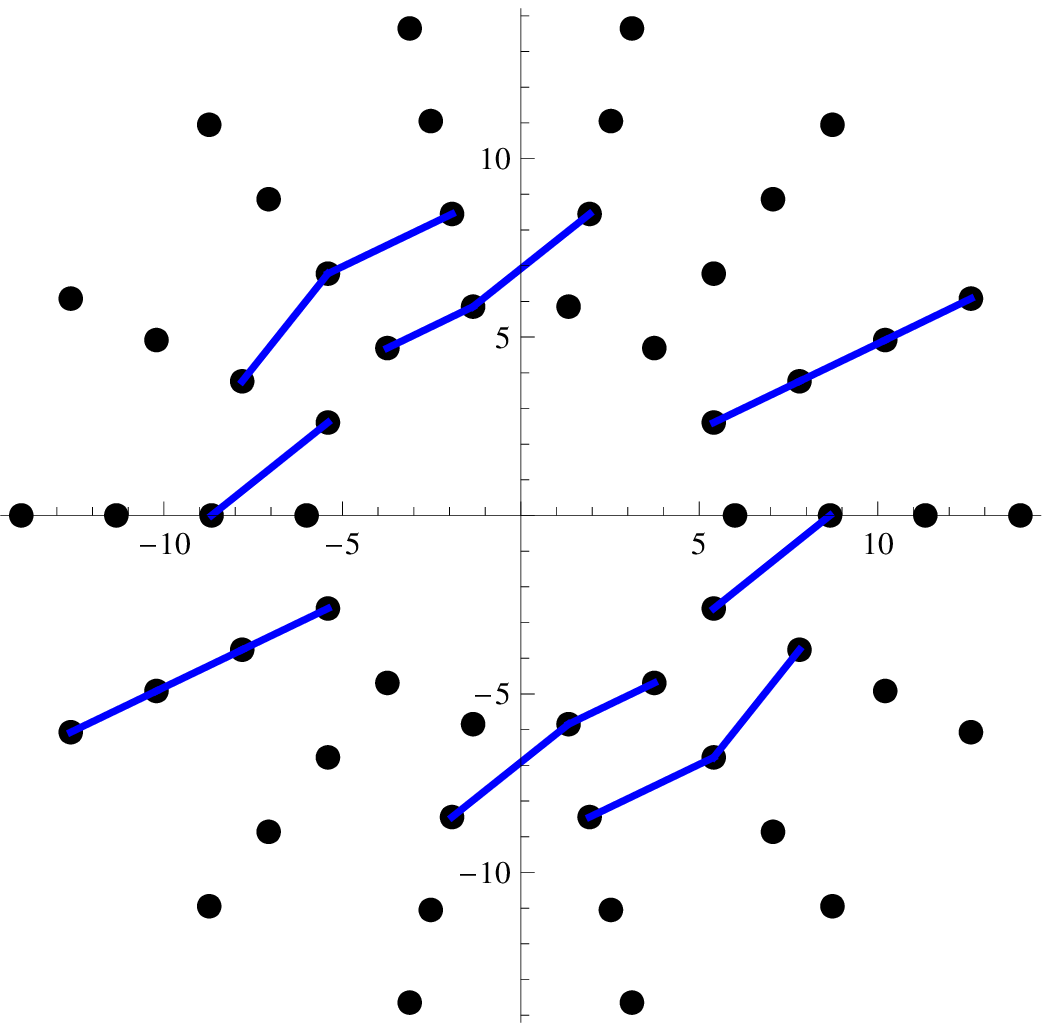}
}\parbox{0.4\hsize}{
\includegraphics[scale=0.5]{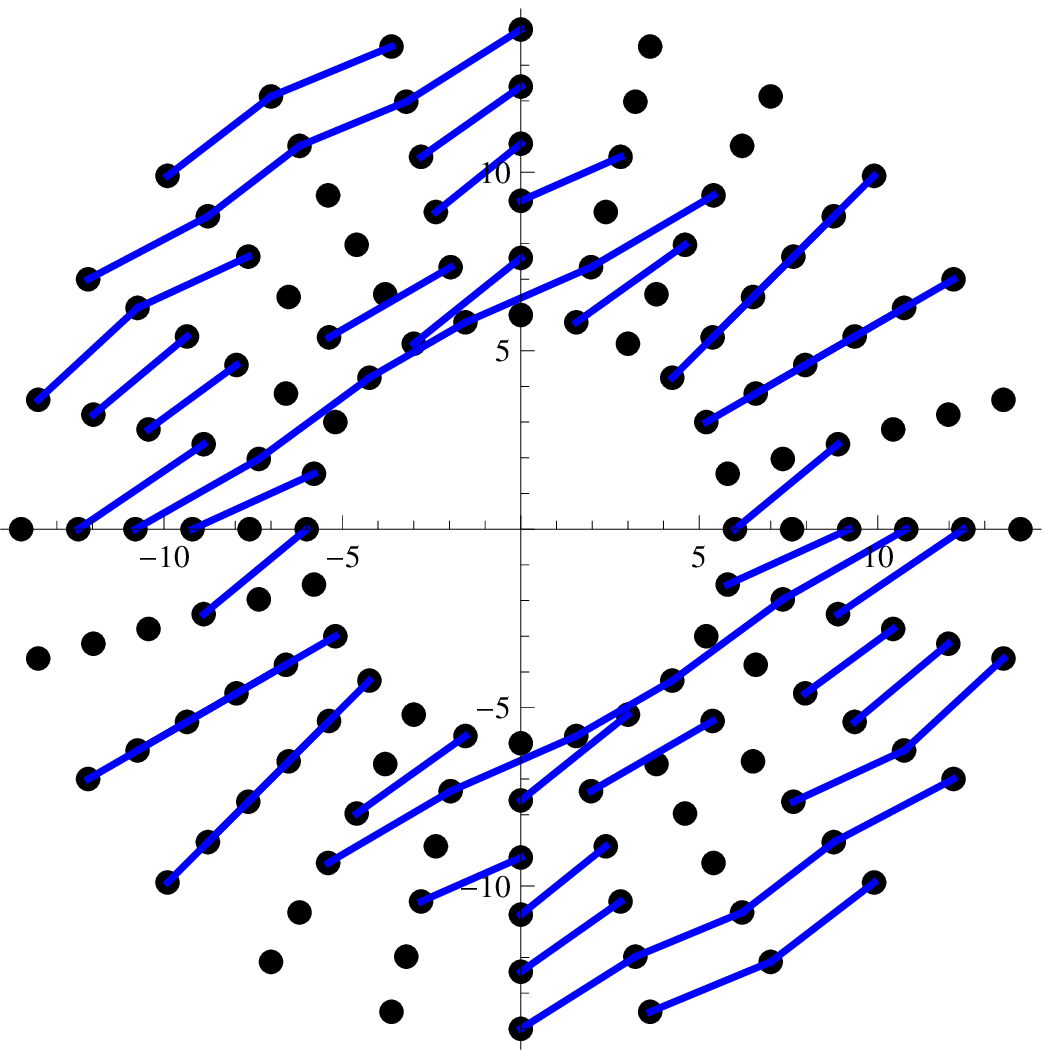}
}\\[5ex]
\parbox{0.4\hsize}{
\includegraphics[scale=0.5]{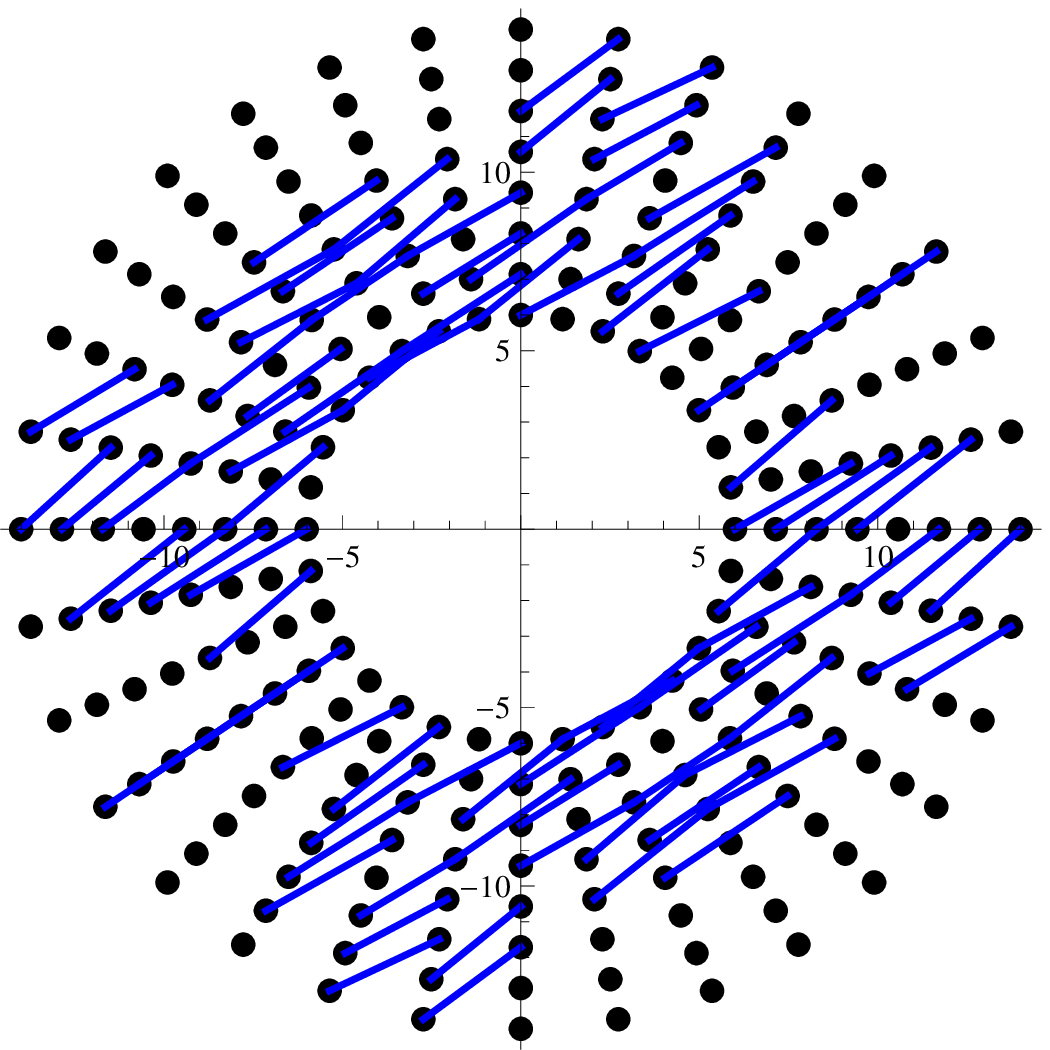}
}\parbox{0.4\hsize}{
\includegraphics[scale=0.5]{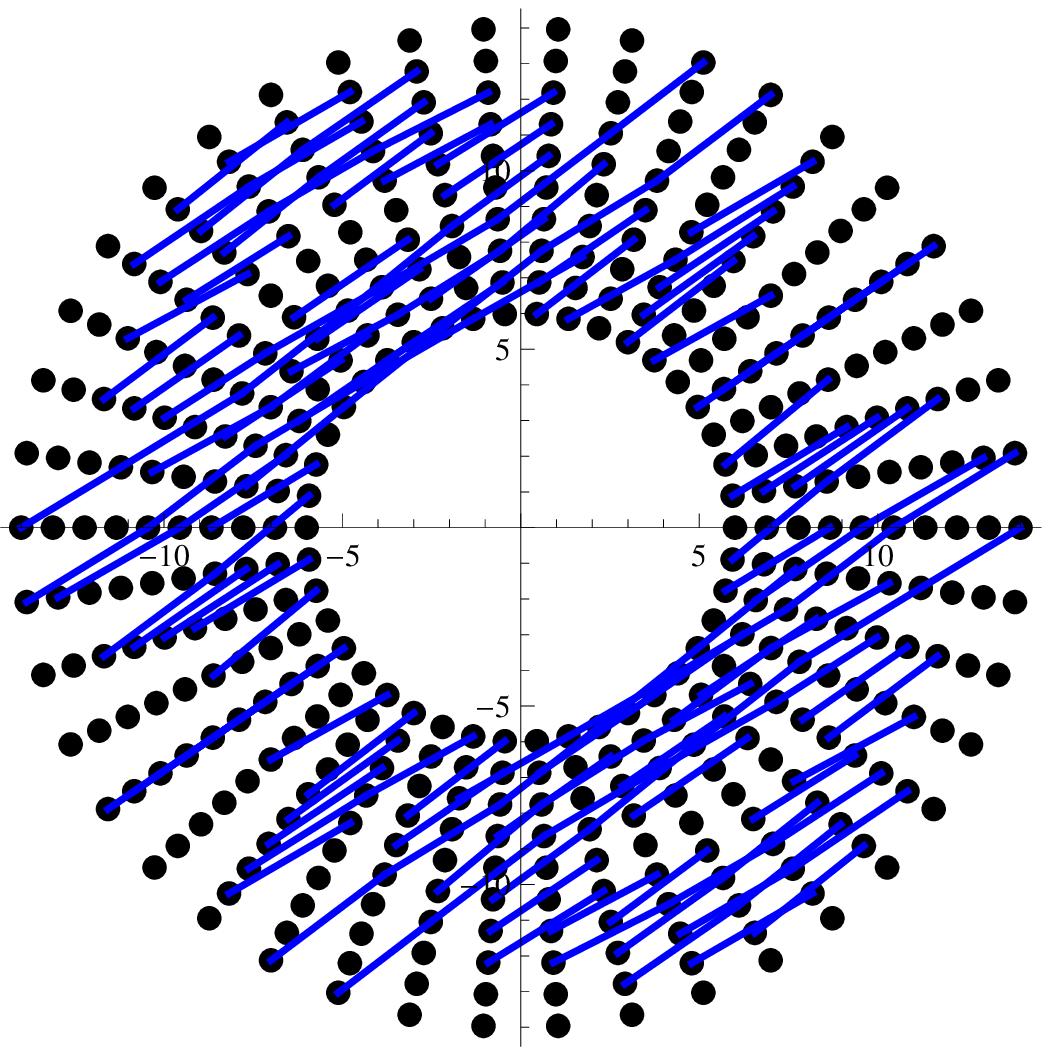}
}
\caption{
Large-volume ($\de p\to 0$) limit showing which pairs of fermion
modes couple to an $A$-mode with momentum $(k_x,k_y)=(3,2)$,
if we tolerate momentum non-conservation $\eps_A=\half\de p$.
We work in units where $p_F=10$ and $\La=4$. The four graphs correspond to
$\de p=2.67$, $\de p=1.6$, $\de p=1.14$, and $\de p= 0.889$.
}
\label{fig:IR-limit}
\end{figure}

We suggest that this problem can be solved by allowing a small
tolerance $\eps_A$ for momentum
mismatch: we will allow a given $A$ mode of momentum $\vec k$ to couple to
any pair of fermionic modes whose momenta differ by an amount $\De \vec p$
that is within $\eps_A$ of $\vec k$, i.e.~$|\De\vec p - \vec k|<\eps_A$.
For simplicity, we will choose the $A$-modes
to live on a cubic lattice with the same momentum lattice spacing $\de p$
as the fermionic lattice, and we will set the tolerance $\eps_A=\half\de p$,
so that typically a pair of fermionic modes will couple to
at most one $A$-mode. The $A$-mode momentum lattice will have
a UV cutoff of $2(p_F+\La)$, so that
fermion modes can be scattered between distant parts of the Fermi surface.
We expect that
in the limit $\de p\to 0$, infrared artifacts
(arising from the momentum lattice spacing and the 
tolerance for momentum non-conservation) will disappear.
In Fig.~\ref{fig:IR-limit} we show how this scheme solves
the problem illustrated in Fig.~\ref{fig:continuum}.
In Fig.~\ref{fig:IR-limit}, as $\de p\to 0$, a given $A$-mode,
in this case the one with momentum $(k_x,k_y)=(3,2)$, couples
to more and more pairs of fermion modes, approximating the
dense set of Fig.~\ref{fig:continuum}a.

In this approach, then, the discretized coupling of the $A$ field to the 
spin-up fermions is given by
\beq
M^{\rm int}_{\up\,p,q,\hat u,\hat v} =
\sum_{\vec k} i A(t,\vec k)
W(p_F({\hat u}-{\hat v})+p{\hat u}-q{\hat v}-\vec k)
\label{Mint-disc}
\eeq
where $W$ imposes momentum conservation with a tolerance of $\half\de p$,
\beq
W(\vec s) = \left\{
 \ba{ll} 1, & \mbox{for}~|\vec s\/|<\half\de p \\[1ex]
         0, & \mbox{for}~|\vec s\/|>\half\de p
 \ea \right.
\eeq

\begin{figure}
\parbox{0.45\hsize}{
\bc
  \includegraphics[scale=0.6]{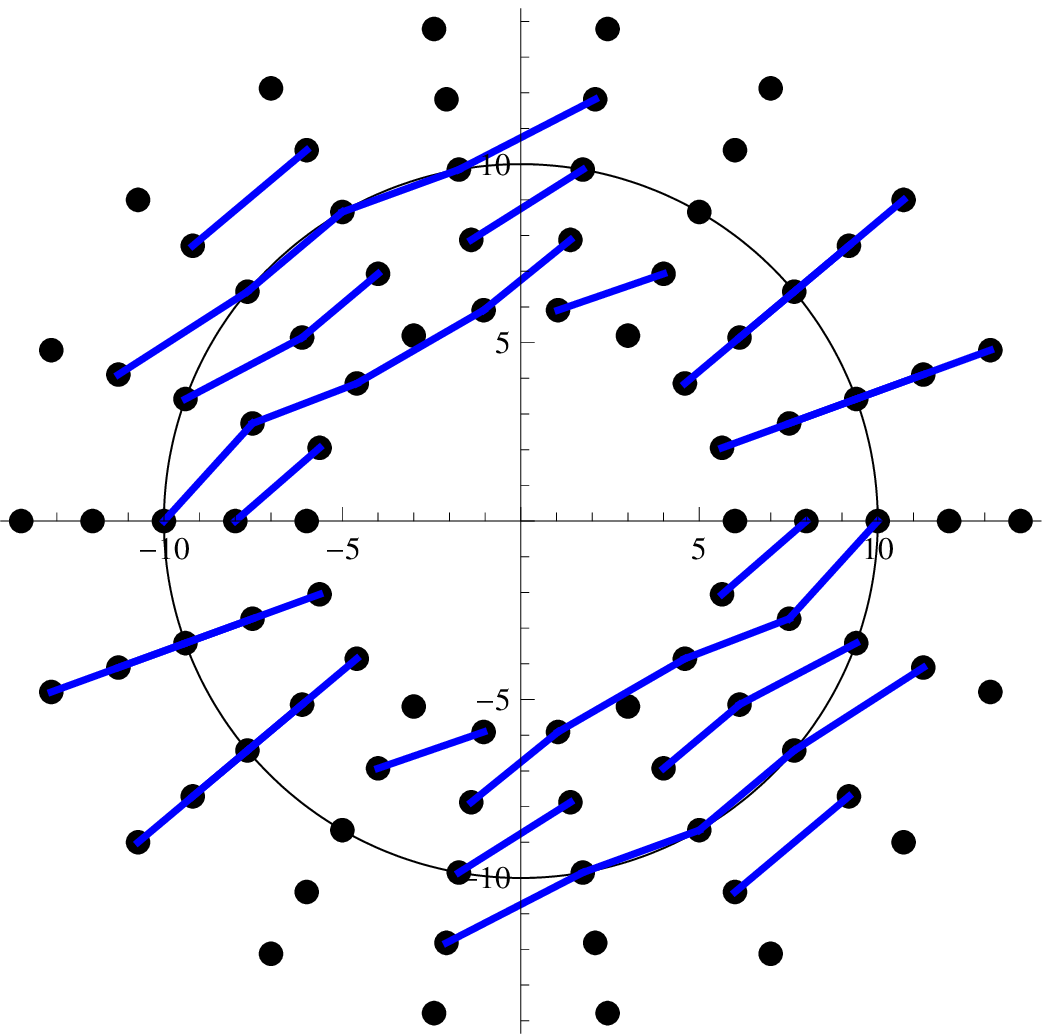}\\
  (a)
\ec}\hspace{0.05\hsize}\parbox{0.45\hsize}{
\bc
  \includegraphics[scale=0.6]{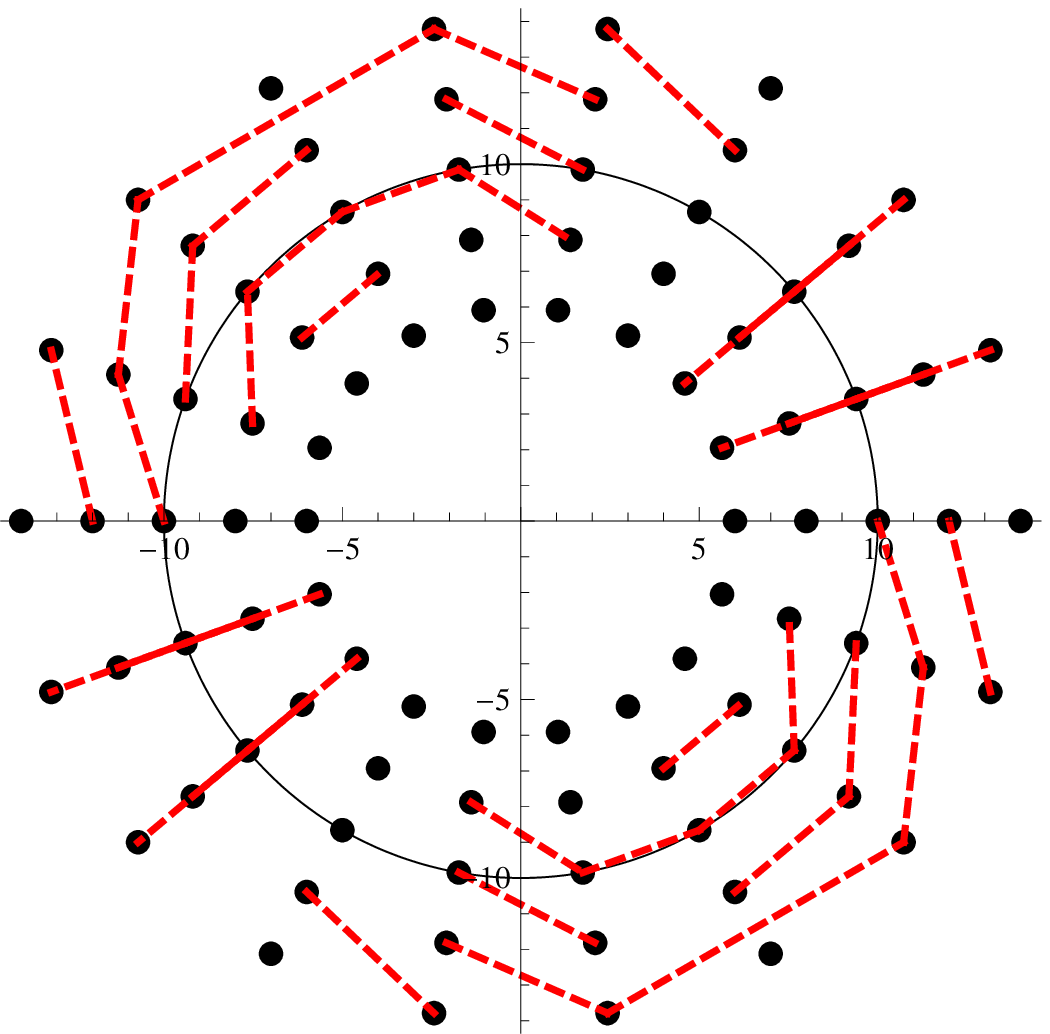}\\
  (b)
\ec}
\caption{Figure showing which pairs of fermion modes
couple to a the $A$-field mode with momentum $(k_x,k_y)=(3,2)$,
for $p_F=10$, UV cutoff $\La=4$, IR cutoff $\de p=2$. 
Panel (a): unmodified theory (see \eqn{M1disc}); momentum violation
is of order $\de p$.
Panel (b): theory with interaction symmetrized in $p$ (see \eqn{Asym},\eqn{M1s})
to ensure a positive semidefinite fermion determinant; momentum violation
is of order $\La$.
}
\label{fig:A-modes}
\end{figure}

In section \ref{sec:discretization} we discussed how the coupling
of the fermions to the auxiliary field could be modified (symmetrized
in the residual momentum $p$) to yield a positive semidefinite fermion matrix.
In Fig.~\ref{fig:A-modes} we illustrate this modification for one of
the auxiliary field modes. We see that in the modified theory 
(Fig.~\ref{fig:A-modes}b) the $A$-mode with momentum $(k_x,k_y)=(3,2)$
couples not only to pairs of fermion modes whose momentum difference
is within about $\de p$ of $k$, but also
to pairs of fermion modes whose momentum
difference deviates from $k$ by up to $\pm\La$. This additional momentum
non-conservation should be corrected by the reweighting back to
the original fermion matrix, and it remains to be seen whether it will
lead to a significant sign problem in computing the reweighting factor.

\subsection{Small antipodal patches}
\label{sec:1patch}

\begin{figure}
\includegraphics[scale=0.4]{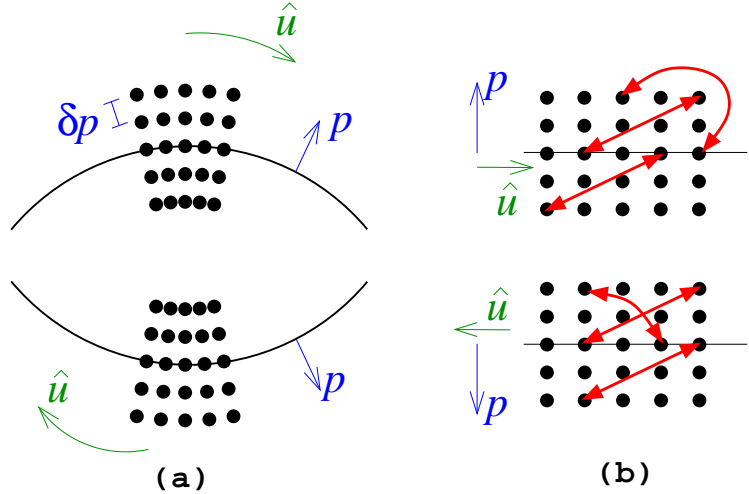}
\caption{Momentum-space discretization of modes in 
two antipodal small patches on the Fermi surface.
Panel (a) shows the modes in polar co-ordinates; panel (b) shows the same
set of modes realized as a
flattened Fermi surface, in which $\hat u$ is a 
periodic Cartesian coordinate orthogonal to $p$.
The double-headed arrows (red online) show some of the 
couplings of the $A$-modes with momentum $(k_x,k_y)=(3,2)\de p$.
}
\label{fig:patch}
\end{figure}

A less complicated coupling of the $A$-modes to the fermions is possible
if we restrict ourselves to a pair of
small antipodal patches on the Fermi surface.
In Fig.~\ref{fig:patch}(a) we show how this
would work in two spatial
dimensions, where $p$ gives
the radial momentum relative to the perturbative Fermi surface, and
$\hat u$ is the polar angle of the momentum.

The approximation of Eq.~\eqn{M1} neglects the difference between
the amount of phase space above and below the Fermi surface, and
can therefore be envisaged as a flattening of the Fermi surface to
a line, as shown in Fig.~\ref{fig:patch}(b), where
$\hat u$ becomes a Cartesian momentum co-ordinate orthogonal to 
the radial momentum $p$.
In each patch we treat $u$ as a periodic variable. 
For simplicity we assume that the lattice spacing has the same 
value $\de p$ in both angular and radial
directions, but one could use an anisotropic lattice
if necessary.
The generalization to three dimensions is straightforward:
simply add another angular variable.

In Eq.~\eqn{M1}, each mode of the $A$-field couples 
equally to all pairs
of fermion modes whose momenta differ by exactly the momentum of
the $A$ field. We implement this in the lattice theory {\em after}
the flattening of the Fermi surface, so the $A$ field lives on
a similar momentum lattice, with the same lattice spacings,
generated by taking all possible
momentum differences on the lattice of Fig.~\ref{fig:patch}(b).
Following Eq.~\eqn{M1}, we allow
each mode of the $A$-field to couple to all pairs
of fermion modes with the right momentum difference.
In Fig.~\ref{fig:patch}(b) we show some of the pairs that one
specific $A$-mode would couple to (double-headed arrows). 
When one takes in to account the
periodic boundary condition in the $u$ direction, these all have the 
same momentum difference.

The small patch scheme is a radical truncation of the full theory, because
it only keeps a small part of the Fermi surface. One could regard 
it as a crude way of implementing a forward-scattering-dominated
interaction. The ground state of a single-patch theory is
more likely to be a particle-hole condensate 
\cite{hep-ph/0307074,Deryagin:1992rw} than a condensate
of Cooper pairs.
But it could also be used as a tractable toy model
for testing the ideas that we have outlined in this article. It is
simpler than the ``whole Fermi surface'' scheme of the
previous section because there is a smaller number of degrees of freedom,
and momentum is exactly conserved.

\subsection{Positivity tests}
\label{sec:tests}

\begin{figure}[h]
\includegraphics[scale=0.6]{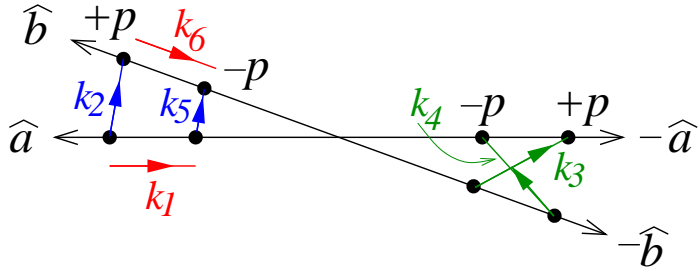}
\caption{
Simplified discretization including only fermion modes with momenta
$\pm(p_F\pm p)\ah$ and $\pm(p_F\pm p)\bh$. These couple to six auxiliary field
modes with momenta $\vk_1\ldots \vk_6$. 
We have shown $\vk_3$ and $\vk_4$ in the antipodal
patches to minimize clutter.
}
\label{fig:antipodal}
\end{figure}

We can test the positivity of the low energy effective theory
in a very simple example, such as that depicted in Fig.~\ref{fig:antipodal}.
We focus on one spin component of the field, and
reduce the fermionic field to two patches, $\ah$ and $\bh$, and their
antipodal partners. In each patch we have two residual momenta $\pm p$,
spaced equally above and below the Fermi surface.
Even if we decouple high-momentum $A$-field modes,
so that there is no coupling between the $(\ah,\bh)$ patches
and the $(-\ah,-\bh)$ patches, the fermion determinant is still not
positive definite. To see this, note that the fermion matrix
is block diagonal
\beq
M^{(\up)} = \left(\ba{cc} M_+ & 0 \\ 0 & M_- \ea\right)
\eeq
where $M_+$ is the fermion matrix for the $(\ah,\bh)$ patches and
$M_- = (M_+)^T$ (see below) is the fermion matrix for the $(-\ah,-\bh)$ patches.
The full fermion determinant is therefore
\beq
\det(M) = \det(M^{(\up)})\det(M^{(\dn)})=\det(M^{(\up)})^2 \ , \qquad
\det(M^{(\up)}) = \det(M_+)^2 \ .
\label{detM}
\eeq
In the basis $(\psi(\ah,+p),\psi(\ah,-p),\psi(b,+p),\psi(b,-p))$,
\beq
M_+ = {\sf P} + i{\sf A}
= \left(\ba{cccc} 
  p & iA(\vk_1) & iA(\vk_2) & iA(\vk_3)\\
  iA(-\vk_1) & -p & iA(\vk_4) & iA(\vk_5)\\
  iA(-\vk_2) & iA(-\vk_3) & p & iA(\vk_6)\\
  iA(-\vk_3) & iA(-\vk_5) & iA(-\vk_6) & -p
\ea\right)
\label{Mplus}
\eeq
where, as shown in Fig.~\ref{fig:antipodal}, $k_1=-2p\ah$,
$k_2=(p_F+p)(\bh-\ah)$, etc. All the $\vk_i$ are much less than $p_F$.
The residual momentum term, {\sf P}, is a diagonal Hermitian matrix 
with entries $\pm p$, and
{\sf A} is an off-diagonal Hermitian matrix because
$A(-\vk)=A(\vk)^*$. We can see that 
$M_- = {\sf P} +i{\sf A}^* = M_+^T$ because $A$ field momenta are reversed
when we perform the transformation to the antipodal momenta.
It is easy to verify that for a generic $A$-field configuration,
$\det(M_+)$ is a generic complex number, so the full fermion determinant
\eqn{detM} of the effective theory is not real.
If we turn off the coupling between neighboring patches by setting
$A(\vk_2)=A(\vk_3)=A(\vk_4)=A(\vk_5)=0$ then the determinant is 
always real, in agreement with the argument of Sec.~\ref{sec:HH}.

\section{Testing using a two-dimensional example}
\label{app:testing}

In this appendix we describe a simple theory that can be used to
test the ideas set forth in this paper. We start with fermions
in two spatial dimensions, feeling a point-like repulsion
\ber
{\rm H}_{int}=
g~\psi^{\dagger}_{\uparrow}({\bf x})\psi_{\uparrow}({\bf x})\psi^{\dagger}_{\downarrow}({\bf x})\psi_{\downarrow}({\bf x}) \ ,
\label{Hint}
\eer
where $g>0$.
We employ Hirsch's variant of the Hubbard-Stratonovich transformation, where
the auxiliary field $\si$ is discrete, taking on values $\pm 1$.
From the start we assume the Euclidean time direction is
periodic with period $T$, and discretized with time spacing $\tau$. 
Then we find \cite{Hirsch:1983je}
\ber
{\rm exp}[-\tau g\,n_\uparrow(\vec{x})\,n_\downarrow(\vec{x})] = \frac12 \sum_{\sigma(\vec{x})=\pm 1}{\rm exp}[2\,b\,\sigma(\vec{x})(n_\uparrow(\vec{x}) - n_\downarrow(\vec{x}))-\frac{\tau g}{2}(n_\uparrow(\vec{x})+n_\downarrow(\vec{x}))]
\label{Hintequality}
\eer
where 
$n_{\alpha}(\vec{x})=\psi^{\dagger}_{\alpha}(\vec{x})\psi_{\alpha}(\vec{x})$,
and $\alpha=\{\uparrow,\downarrow\}$, and
\ber
\tilde b= \frac {b}{\tau} \ ,
 \qquad {\rm tanh}^2(b)={\rm tanh}\left(\frac{\tau g}{4}\right) \ .
\label{b}
\eer

The fermion matrices in the effective theory in momentum space
are then given by
\ber
&&{\rm M}_{F\uparrow}=\left[\partial_t+\frac{p_F}{m}~p\right]\delta_{{p},{q}}\delta_{{\hat u},{\hat v}}+2\,{\tilde b}\,\sigma(t,p_F({\hat u}-{\hat v})+p{\hat u}-q{\hat v}),\nonumber \\
&&{\rm M}_{F\downarrow}=\left[\partial_t-\frac{p_F}{m}~p\right]\delta_{{p},{q}}\delta_{{\hat u},{\hat v}}+2\,{\tilde b}\,\sigma(t,-p_F({\hat u}-{\hat v})-p{\hat u}+q{\hat v}),
\label{MF}
\eer
where $p_F=\sqrt{2 m \bar{\mu}}$ and ${\bar{\mu}}=\mu-{g}/{2}$ as follows from (\ref{Hintequality}). Note that we use a Nambu-Gor'kov-like basis where
the $\sigma(t,{\bf p}-{\bf q})$ term is positive
in both the spin-up and spin-down fermion matrices. The auxiliary field
does not have any action of its own (${\rm S}_{aux}=0$). 

We now discretize the system for lattice calculations. 
The radial residual momentum has an infra-red cutoff $\de p$
and ultra-violet cutoff $\La$, and the angular variable
is discretized in $N_v$ steps
(see Sec.~\ref{sec:discretization})
so the momentum lattice action is
\beq
\ba{rl}
\dsp {\rm S}_{lat}=\frac{p_F L}{N_v}
 \!\!\sum_{k,n=-N_p}^{N_p}\sum_{u,v=0}^{N_v-1}
 &\dsp \psi^{\dagger}_{{u},p_k}\Bigl[\delta_{u,v}\delta_{k,n}\left(\nabla_t+\frac{p_F p_n}{m}\right) \\[2ex]
 &\dsp +\,2\,{\tilde b}\frac{p_F L}{N_v}\sigma(p_F({\hat u}-{\hat v})+p_k{\hat u}-p_n{\hat v})\Bigr]\psi_{{v},p_n}.
\ea
\label{Slat}
\eeq
where $N_p=\La/\de p$, $p_n = n \de p$, $\hat v$ is a unit vector at 
angle $2\pi v/N_v$ to the $x$-axis, we have defined the IR cutoff 
length scale $L=2\pi/\de p$, and $\nabla_t$ is a discretized
version of the time derivative.
The fermion matrix dimensionality is $N_t\times N_v\times(2 N_p+1)$.


The auxiliary fields, $\sigma(t,{\bf p}),$ are defined on a square lattice
in Cartesian momentum space, with UV cutoffs $\pm \Lambda_\si$ on the
$x$ and $y$ components of momentum, where $\Lambda_\si=p_F+\Lambda$, and
a momentum lattice spacing (IR cutoff) $\de p_\si$, and we define
$L_\si=2\pi/\de p_\si$. In practice we work with a position space
square lattice on which $\sigma(t,{\bf x})=\pm 1$ and we Fourier transform
it to obtain  $\sigma(t,{\bf p})$. We set $\de p_\si=\de p$.
The number of sites along
each side of the $\si$-lattice is an odd integer, $2\La_\si/\de p_\si + 1$.

We have performed very preliminary calculations for
$p_F=2.58$, $m=1.0$, $\bar\mu=3.33$, $T=1.0$, with coupling
$g=1.0$.
For the fermions we used a momentum space lattice with $\de p=0.194$, 
$\La=0.775$, $N_v=64$. This corresponds to $N_p=4$,
$\La/p_F=0.3$, $T/{\bar\mu}=0.3$.
In the Euclidean time direction we set $\tau=0.111$ which corresponds to 
$N_t=9$. The dimension of the fermion matrix is $5184$.
For the auxiliary field we use $\La_\si=p_F+\La=3.35$
$\de p_\si=\de p=0.194$, so
the $\si$ field lives on a $35\times 35$ cubic lattice.


To get an idea of the severity of the sign problem in this formulation
of the theory, we performed a Monte-Carlo calculation of the
reweighting factor ${\cal R}$ (see Sec.~\ref{sec:conclusions}),
We formulated the theory in {\em Mathematica}
using its {\sf Det}
function to numerically evaluate the determinant of the 
$5184\times 5184$ fermion matrix.
We used the Metropolis algorithm, with an accept/reject step for
each update of the auxiliary field $\si$ using the non-negative
weighting factor $W_{\rm pos}=|\det {\rm M}_{F\uparrow}{\rm M}_{F\downarrow}|$.
We note that even with this numerically unsophisticated approach, we are
able to study a lattice that covers an entire 2D Fermi surface, with
reasonable resolution ($\La$ four times bigger than $\de p$ \eqn{lat_ineq}).

The observable that we measured is the real part of the
reweighting factor, 
${\rm Re}{\cal R} = \cos(
 {\rm arg}\det {\rm M}_{F\uparrow}{\rm M}_{F\downarrow}(\sigma)
)$.
In our Monte-Carlo runs we started with all the $\si(t,{\bf x})$ 
being $+1$, and found that when we had done enough updates
(about $10^4$)
the $\si$ were equally distributed between $+1$ and $-1$, and
the average value of ${\rm Re}{\cal R}(\sigma)-1$ was of order $10^{-8}$,
with occasional fluctuations no larger than about $10^{-2}$.
These initial results have not been checked for proper
thermalization, autocorrelation, and finite volume effects, but they
are an encouraging indication that the theory of the degrees of
freedom near the Fermi surface has a moderate sign problem, with a
reweighting factor that is close to 1, rather than fluctuating around
zero.

\renewcommand{\href}[2]{#2}

\bibliographystyle{JHEP_MGA}
\bibliography{sign_problem} 

\end{document}